# HUMAN NICHE EVOLUTION: PATHWAYS, CHOICES AND OUTCOMES

Miguel da Silva Pinheiro [1], Pablo José Francisco Pena Rodrigues[1]

[1]Instituto de Pesquisas Jardim Botânico do Rio de Janeiro, Rua Pacheco Leão, 915, CEP 22460-030, Rio de Janeiro, Brazil

**\*Corresponding author:** Rodrigues, P.J.F.P. (pablojfpr@hotmail.com)

**Abstract**

Humankind has spread worldwide supported by cultural and technological knowledge, but the environmental sustainability on the human niche evolution depends on a new human beings relationship with the biosphere. Human lifestyles nowadays are very "Antropocentric" and in many ways deleterious to the other life forms. Here we try to identify future scenarios, where the less deleterious is the "Natural-Technological Model" that points the urgent need to change the evolutionary direction of the human niche seeking the resumption of original ecological relations. New cultural habits and novel technologies, thereby, would reverse the current anthropogenic impacts. The middle way is the Bio-Anthropogenic Model that predicts the success of the emerging ecosystems and the symbiotic relationship of humans and anthropogenic-favored species, hybrids, aliens and genetically modified organisms. For such, we must also change our way of life and adopt new conscious ways of consumption aiming at the socio-environmental good. Lastly, the Wear Out Model, which depends only on maintaining current patterns of human expansion. The lack of investments on new technologies and new cultural habits, added to the current patterns of human niche evolution that are based on the massive exploitation of world resources, will lead to a fearsome scenario with a precarious global health, biodiversity losses and food scarcity. This theoretical models indicates some pathways and can help us to choose a better future.

**Keywords: Environment; Restoration; Anthropogenic; Sustainability; Collapse**

**Resumo**

A humanidade se espalhou pelo mundo apoiada no conhecimento cultural e tecnológico, mas a sustentabilidade ambiental na evolução do nicho humano depende de uma nova relação do ser humano com a biosfera. O estilo de vida humano hoje em dia é muito "Antropocêntrico" e, em muitos aspectos, deletério para as outras formas de vida. Procuramos aqui identificar cenários futuros, onde o menos deletério é o "Modelo Natural-Tecnológico" que aponta a necessidade urgente de mudar o rumo evolutivo do nicho humano buscando a retomada das relações ecológicas originais. Novos hábitos culturais e novas tecnologias, assim, reverteriam os atuais impactos antrópicos. O caminho do meio é o modelo bio-antropogênico que prevê o sucesso dos ecossistemas emergentes e a relação simbiótica entre humanos e espécies antropogênicas favorecidas, híbridos, alienígenas e organismos geneticamente modificados. Para tanto, devemos também mudar o nosso modo de vida e adotar novas formas de consumo consciente visando o bem socioambiental. Por fim, o Modelo de Desgaste, que depende apenas da manutenção dos padrões atuais de expansão humana. A falta de investimentos em novas tecnologias e novos hábitos culturais, somados aos atuais padrões de evolução de nicho humano que se baseiam na exploração massiva dos recursos mundiais, levarão a um cenário assustador de precariedade global da saúde, perdas de biodiversidade e escassez de alimentos. Esses modelos teóricos indicam alguns caminhos e podem nos ajudar a escolher um futuro melhor.

**Introduction**

Ecosystems have a great temporal change capacity (Engel et al., 2004), therefore they are dynamic and undergo relevant fluctuations over time and space. These changes occur due to short, medium, and long-term environmental changes (Kimmins, 2004), which can have natural or even anthropogenic causes (Rosenzwieg et al., 2008). The changes bring new environmental dynamics, enabling a moment of reorganization for Ecology. Currently, one of the main causes of these changes are humans (Albuquerque et al., 2018) and their niche in the planet. This impact on Earth is unquestionable, showing itself present in the extinctions of the Quaternary Megafauna and even in the complexity of industrial economies, expanding populations and dense transport networks of contemporary human societies (Boivin et al., 2016). In view of these massive habitat changes, the human beings would not only change his own evolutionary path, leading to great adaptive changes in other animal and plant species, acting intensely as a modulator of natural and artificial selection, whether intentional or not (Sullivan et al., 2017).

One of the main consequences of the history of human occupation is environmental degradation, yet there are currently many attempts to reverse this process through "Ecological Restoration" that have great potential to minimize human impacts on the biosphere. However, degradation occurs at multiple scales of space and time and among the main factors are the emission of mass pollutants, ranging from synthetic materials to heavy metals, whose presence in organisms affects their physiological systems and can result in serious diseases, contaminating trophic networks, often affecting humans (Rhind, 2009).

Within the scope of Ecological Restoration, several approaches and techniques have been developed to achieve the restoration of ecosystems, causing discussions about which methods have the best results (Meli et al., 2017). However, by targeting certain so-called "ideal" scenarios for the restoration of a given environment, we often end up interfering even more in its balance, contributing to its decline (Hobbs et al., 2011). It is then necessary to restructure the concepts of environmental restoration on a global basis, aiming to mitigate the phenomena of environmental degradation and climate change, providing paths for a social-environmental balance. This work seeks to improve the theoretical framework and propose new perspectives about the concepts of human niche evolution and ecological restoration. This, aiming at the environmental balance, that may result in an increase in the quality of life for both other beings and *Homo sapiens* itself.

### Evolution of the human niche and ecosystem changes

The debate around the human niche expansion and how it has been affecting the diverse ecosystems around the globe is growing. To understand these impacts it is necessary to know the human nature in depth, considering their differentiated psychological and social characteristics, to generate sets of interpretative knowledge, through records of our evolutionary past until today. An extremely important factor in recognizing our niche would be that it is based not only on the evolutionary strategy of advanced cognition and genetic inheritance, but also on the ability of *Homo sapiens* to develop cultures, often thousands of years old, and to shape themselves according to these, taking a unique evolutionary pattern process (Fuentes, 2017).

The genus Homo, during the Pleistocene, takes an evolutionary path regarding the increase of brain mass and its complexity of functioning, developing concepts of complex social structures (Van Schaik, 2016). With these acquired characteristics, the enhancement of the absorption of knowledge and resilience to the adversities that occurred, in which one of the greatest differentials of the expansion of the human niche was the use of extra-somatic materials in innovative ways (Fuentes, 2016).This complex and distinctive character of the human niche has several consequences due to its influence on the ecology of different communities throughout history, owing to the peculiarities of human ecosystem engineering (Boivin et al., 2016). In view of the massive modulations of the environment, humans end up changing their own evolutionary path (Rodrigues & Lira, 2019), in addition to creating scenarios conducive to considerable adaptive changes in other animal and plant species, acting intensely as a modulator of artificial selection (Sullivan et al., 2017). This great dominance in addition to constituting new evolutionary pathways, also contributes to the creation of unique organisms, establishing man as a "hyperkeystone species" (Worm and Paine 2016).

The occupation of a distinct and unique niche is highlighted in humans, considering the different evolutionary narratives of our species, in addition to the nuances of the sophisticated cognitive aspect of the genus Homo (Fuentes, 2017). Therefore, the main differences of the human niche (Ellis et al., 2016) for the model common to all organisms included in the spatial, ecological and social pillars, is our capacity to synthesize contexts of perceptual patterns through extremely complex manipulation of the ecology around us (Boivin et al., 2016), in which we impose selective pressures and adaptive advantages on a huge range of species with which we share the environment (Sullivan et al., 2017), often introduced by the questions and modalities attributed to human cultures (Fuentes, 2017).

By recognizing humans as niche creators and extremely influential ecosystem engineers (Albuquerque et al., 2018), we identify the major changes we cause on the biodiversity, including: extinctions, extirpations, changes in the composition of species, in addition to changes in their diversity and community structures. These events also coincide with four main phases: the expansion of human occupation in the late Pleistocene, as wells as the expansion of agriculture in the Neolithic, the process of colonization of islands and the emergence of the first urbanized societies and commercial networks, both in the Anthropocene (Boivin et al., 2016).

**Anthropogenic environmental impacts and collapses**

Among the main anthropogenic modifying actions across the globe is the extinction of the megafauna in the late Quaternary, in which large herbivores were hunted to extinction by primitive groups of *Homo sapiens* (Surovell et al., 2005; Johnson, 2006; Koch & Barnosky, 2006). These extinctions resulted in several environmental changes, altering various biotic and abiotic factors, such as water availability and rainfall. As a result, several habitats such as the steppes have disappeared around the globe and several plant species have become extinct (Johnson, 2009).

The extinctions that currently occur are similar to what occurred in the Quaternary period and scenarios of better conservation and resource management could be achieved if efforts were directed at the main causes. However, the post-industrial revolution scenarios show that the degradation was increasing and the exploration and search for resources ended up damaging the environment, generating several planetary consequences, such as global warming, acidification of the oceans and depletion of resources, in addition to the mass emission of pollutants (Rosa et al., 2004).

The depletion of natural resources through predatory fishing by industries has been causing the collapse of coastal ecosystems (Jackson et al., 2001), especially where there is inadequate management. Therefore, changes in water quality and the occurrence of several cascading impacts on trophic networks are highlighted, due to explosions in phytoplankton communities to the detriment of fish populations (Daskalov, 2002). There are forecasts in which, starting from the middle of the 21st century, fishing resources globally will collapse, irreversibly (WWF, 2015).

The Human beings are causing environmental imbalances to all species and for himself. A clear pattern of this appears when we observe food chain contamination and its ecological impacts on the various participating populations (Rhind, 2009), including the human population. Due to the large emissions of pollutants and micro-plastics in the oceans, only 8 million tons of them are released annually (Jambeck et al., 2015). Thus, much of the food we consume from the marine environment is contaminated. In addition, the excessive consumption of these pollutants can affect human health, ranging from skin irritations, decreased fertility rate to the occurrence of cancers (Carbery et al., 2018).

However, despite all the anthropogenic impacts and environmental stresses experienced in the past and today, most of the environmental functions can be restored through less destructive ways of exploring the environment. An illustration of this reversal of frames previously seen as critical would be given during the current pandemic of COVID-19, in which several countries, such as Italy and India, assessed air quality in several locations considered problematic after a few months of mandatory lockdown, revealing a significant increase in the quality of the air available in these regions (Rugani & Caro, 2020; Singh & Chauhan, 2020). In addition, the technological advances provided by the post-industrial revolution scenarios present us with a range of tools aimed at environmental preservation, allowing the restoration of ecosystems more quickly and efficiently. The way human beings deal with resource management, today, is linked to consumerism and unnecessary waste (Sehrawat et al., 2015), and it is fundamental to reshape the perspectives and ideals of human consumption aiming at sustainable results in different situations (Meroni, 2007), such as the so-called "green consumerism", as well as the use of so-called green technologies (Krass et al., 2013).

**Ecological Restoration and its perspectives**

In view of the current interpretations of Ecological Restoration, the most used concepts today would have mainly a "modifying" character towards the target ecosystem (Hobbs et al., 2011). In this sense, the restoration would aim to promote the return of original interspecific relationships, in addition to the presence and absence of certain species such as to a previous or even pristine state. Completely modifying the new conditions currently imposed by humanity. In addition, the conceptual basis for the return of the environment referring to this "original state" would be difficult to establish, since the conditions and information for such a restructuring model are often non-existent, making it impossible to return to a pristine stage of environmental balance.

To reduce this infeasibility of ecosystem restoration, Ecological Restoration should undergo a certain restructuring, aiming at the use and exploitation of the maintenance and balance of current ecosystems (e.g., Arroyo-Rodríguez et al., 2015). Assisting the aspects of ecological relationships currently established in conjunction with their evolution, aiming to advance the succession processes of ecosystems, often classified as novel (Hobbs et al., 2006) or anthropogenic (Ellis et al., 2008), increasing their biodiversity and helping to balance it in future scenarios.

Within the scope of ecological restoration in ecosystems, the methods presented are usually effective and help in the recovery of biodiversity and maintenance of ecosystem services in these landscapes (Costanza et al., 1997). The restoration strategy adopted is variable for each case (Barral et al., 2015). When analyzing the effectiveness of restoration methods, in many cases, natural regeneration through passive restoration proves to be effective, as is the case with certain tropical scenarios (Uriarte et al., 2016). However, many variables are considered for the application of these methods, and human interference is often necessary when entering active restoration methods (Meli et al., 2017).

Passive restoration, which is based on natural regeneration, advocates minimal human intervention favoring the recovery of environments in a more spontaneous way, in which native species would naturally colonize abandoned fields after previous natural disturbances, establishing a new community (Shono et al., 2007). Methods within this strategic field can often be aided by anthropogenic intervention, through pest control and protection against possible fires (Zahawi et al., 2014). Meanwhile, active restoration in many cases promotes the planting of seedlings cultivated in nurseries, changes in disturbance regimes related to deforestation and burning, and direct seeding to accelerate the regeneration process (Holl et al., 2011), to improve the species composition or allow a new scenario of ecological succession (Chazdon et al., 2016). Thus, it is evident that strong human interference can have important consequences on the restored communities, such as the establishment of new tree communities distinct from the original vegetation. In other cases, the target communities for restoration are already modified ecosystems, and it is possible to use local dynamics aiming to reduce human interference.

When understanding the dynamics of some degraded areas, it is possible to outline a strategy for the restoration of these sites, taking into consideration: The cost of the operation, the desired results, the residual vegetation, and the level of degradation (Chazdon, 2008). The use of different strategies to solve this complex problem implies knowing different environmental characteristics of the studied place. Applying, for example, the strategy of "vegetation islands", which provide a complex structure with nucleation techniques, acting to facilitate the restoration of space and mimicking a natural occurrence (Yarranton & Morrison, 1974; Zahawi & Augspurger, 2006).

The concepts of stability and resilience derive from the potential for restructuring ecosystems, which express the speed with which the variables of a system return to equilibrium after a disturbance (Côté et al., 2010). The resilience of an ecosystem is often related to the availability of water, light, nutrients, physical properties of the soil and the use of the current soil, since after anthropogenic interventions, invasive vegetation becomes very aggressive in competition for available resources (Elmqvist et al., 2003).

As for the cost involved for the different approaches (Chazdon et al., 2008), using natural regeneration the costs are considerably lower when compared to active restoration (Birch et al., 2010). In addition, recent studies demonstrate the use of passive restoration can generate similar results in plant diversity even when compared to those obtained from active restoration processes (Shoo et al., 2015). Therefore, in scenarios of favorable ecological conditions, natural regeneration seems to be the less expensive approach to recovery of diversity and ecological processes (Poorter et al., 2016).

Thus, it is pertinent to discuss Restoration models that consider the succession processes in natural regeneration, which discards the need for new plantings and makes the restoration process more effective in locations with a high level of aptitude and resilience. Understanding the correlation between restoration and its ecosystem services brings interesting results not only for the science of restoration, but also for improvements in the quality of life of the human population, in the economy and in government policies (Naidoo et al., 2008).

**Bio-evolutionary Anthropocene creating perspectives**

Changes in the Biosphere due to human influence are at the basis of the Bio-Evolutionary Anthropocene theory (Pena Rodrigues & Lira, 2019). The human relationship with different organisms that have had some anthropogenic influence, such as hybrids, genetically modified or aliens, communicates directly with new proposals for restructuring ecological restoration, since through their management and their interactions with other individuals, they create unexpected evolutionary paths and different scenarios. Through the analysis of these different perspectives, the scenario referring to the Bio-Evolutionary Anthropocene would be one of the ways to achieve ecological balance on a global scale. Thus, species adapted to human interventions, that is, favored by this spectrum, would have their density increased, giving them a character close to a "symbiosis" with humans.

On the other hand, the concept of Ecological Heritage refers to legacies of change, both in biotic and abiotic environments, by niche-building organisms, which modify selection pressures on descendant organisms (Odling-Smee et al., 2013). Furthermore, the favoring of some species is considered due to the anthropogenic influence, being possible the occurrence of persistence along a time gradient of pressures and selections dictated by human modifications, considered an extremely influential factor for the theory of niche construction (Bonduriansky & Day, 2009; Danchin et al., 2011; Bonduriansky, 2012), directly integrated in the evolutionary progress of the human niche. In addition, environmental modifications mediated by organisms, especially when the object of study is the Homo sapiens, end up presenting a wide range of natural community evolutionary effects indirectly in several species (Odling-Smee et al., 2013; Walsh, 2013). It is possible to consider a diverse collection of selective pressure agents that could drive evolution, by establishing accumulated phenotypic covariance between organisms, resulting in changes in biotic organisms and abiotic environmental conditions (Matthews et al., 2014).

In the construction of the human niche related to domestication, the initial stage of movement of fauna and flora under human management in pioneer societies, even in a scenario of lack of resource depletion and environmental imbalance, occurred in favorable environments and low population densities (Smith, 2016). These impacts are considered relatively less harmful to the ecology of the new territory occupied by man, compared to the current moment. The impact, said to be less harmful, in a small-scale scenario ends up becoming a global problem, in periods after the industrial revolution, creating situations of extreme habitat changes, due to the replacement of native species by anthropogenically favored species (Pena Rodrigues & Lira, 2019). Therefore, human hyper-dominance is a key factor in the construction of the human niche and in the relationships established with other species.

By analyzing the perspective of the Bio-Evolutionary Anthropocene, it would help us to find new paths to achieve the expected Ecological Restoration, through the recognition of these environmental changes managed by Homo sapiens, aiming at the flourishing of novel (Hobbs et al., 2006) or anthropogenic ecosystems (Ellis et al., 2008) in future scenarios.

**Paths, Choices and Outcomes**

Reassessing the concepts of Human Niche Evolution and Ecological Restoration, we will be able to point out different paths. Through new perspectives we can analyze how our choices will influence and what results we will reach. However, human influence on the environment is dynamic, so a wide range of options for projects aimed at restoration must be considered, each adhering to the peculiarities of the studied environment. It is possible to identify, initially, the paths that would appear before certain attitudes regarding our relationship with the environment, dividing into three main categories: Natural-Technological Model, Bio-Anthropogenic Model and Wear Model.

*The Natural-Technological Model*

The Natural-Technological Model, would aim to change the evolutionary direction of the human niche, aiming at reestablishing ecological relationships between species originating in the studied locations (Hobbs et al., 2011), to reach the balance of these ecosystems, as well as the creation and use of new technologies and ways of life, aiming to significantly mitigate global anthropogenic impacts (Meroni, 2007).

The applicability of changes in the human niche, in this scenario, would be attributed to resilience to adversity, in which the crisis factor would be environmental degradation (Redclift, 2010). The potential for assimilation of human knowledge would be the differential of our specie, since through cultural changes towards the natural environment it would be possible to synthesize significant changes in the customs of future generations, once again using extra-somatic materials in different and innovative ways (Fuentes, 2016), being the key to human evolution.

The discussion of new models of human occupation is necessary (Kelbaugh, 2019), since to achieve this scenario, several choices to optimize the space available for the greater effectiveness of the restoration processes must occur, being aimed mainly at localities and urban centers. From the management of living models and urban technology, it would be possible to adopt architectures similar to Japanese society (Karan, 2005), known as kyosho jutaku, or micro houses, in which the main objective in residential planning would be to improve spatial use in a few square meters, allowing the expansion of the limits of urban forest fragments due to spatial management. It is also appropriate to discuss the human population size, in which policies aimed at birth control could lead to fewer needs for the expansion of human occupation to the detriment of the environment (Speidel et al., 2007).

As for the use of Ecological Restoration in natural areas, a large part of its resources would be aimed at promoting active strategies, aiming at the integral acceleration of the global regeneration process (Holl et al., 2011) and recovery of interspecific interactions originating in the place (Hobbs et al., 2011), especially in areas with low resilience and consequently low aptitude for passive restoration. In addition, the implementation of agroforestry models to the detriment of landowning properties, would be extremely necessary for the advancement of the proposed social-ecological system, since it would allow the economic movement of natural resources together with the aid and maintenance of ecosystem recovery (Noordwijk et al., 2020). Another form of environmental preservation would be through prioritizing the local production of food, related to the micro-economy, since this would reduce the need to purchase products from the large estate agribusiness, in which the generation of commodities are focusing on feeding livestock, which on large scales is extremely harmful to the environment (Steinfeld & Gerber, 2010; Berti & Mulligan, 2016).

Finally, it would be necessary to rethink the uses of technologies aimed at mitigating anthropogenic impacts, especially in relation to the management of resources wasted daily (Sehrawat et al., 2015) and emission of pollutants. The implementation of "green technologies" on a large scale would be ideal for the Natural-Technological Model, as these would aim to significantly reduce the emission of polluting gases and waste, long-term operation, and lower rates of wasted resources, having projects applied significantly in countries like China and India (Lema & Lema, 2012). Being articulated as the reduction of the cost of the so-called sustainable technologies for its implementation on a global scale, adversity resolved with government support in the form of subsidies or reduction of import taxes and other taxes related to the manufacture and use of "green technologies" (Krass et al., 2013).

*Bio-Anthropogenic Model*

In view of the concept of the Bio-Evolutionary Anthropocene (Pena Rodrigues & Lira, 2019), the Bio-Anthropogenic Model is synthesized, basing its viability on the anthropogenic (Ellis et al., 2008) and emerging ecosystems (Hobbs et al., 2006). This scenario would characterize, then, changes in the interpretation of the concept of Ecological Restoration, aiming at the environmental balance through the interspecific interactions currently existing on account of Ecological Heritage (Bonduriansky & Day, 2009; Danchin et al., 2011; Bonduriansky, 2012), being these are often the result of human interference and the evolution of their niche, in which the species favored anthropogenically would have a greater distribution. In addition, it would still be necessary to seek less-impacting forms of development, with the idea of "green consumerism" being extremely applicable to the model (Meroni, 2007).

The most predominant character of the human niche in this scenario would be its peculiar influence as a global ecosystem engineer, calling Homo sapiens a "key hyper-species" (Worm & Paine 2016). The ability of community ecological modulations contributes to the construction of new evolutionary routes and the synthesis of new organisms (Sullivan et al., 2017), with this we have unprecedented ecosystem dynamics, which can demonstrate scenarios of increased resilience to certain species (Hobbs et al., 2009), which can be considered anthropogenic favored.

Passive restoration strategies can be applied to the Bio-Anthropogenic model, since through the recovery of forest environments spontaneously, after anthropic disturbances, a new community dynamic would be established (Shono et al., 2007), synthesizing a new emerging ecosystem (Hallet et al., 2013) adapted to new local conditions. The increase in local resilience would occur precisely because of the consideration of the new interspecific relationships of the habitat (Elmqvist et al., 2003), in which so-called exotic or invasive species would not be attributed to a negative character for ecosystem suitability, as in extremely anthropized ecosystems, the species cited as aliens may present themselves as resistant or even favored by human activities (eg, Milla et al., 2015).

In addition, it is necessary to analyze the favoring of hybrid organisms inserted in this model, since these adapted individuals would present an extremely important form of genetic variation for processes of speciation and evolutionary lines emerging in scenarios of human influence (Barrera-Guzmán et al., 2017). Considering, also, that hybrid organisms demonstrate high evolutionary capacity, evolving rapidly in isolation scenarios of individuals (Schumer et al., 2015), and most of these beings with a history of establishment in natural environments showed greater adaptive capacity than their parental organisms (Crispo et al., 2011).

In the Technosphere, it is worth emphasizing the implementation of genetically modified organisms, both as measures to ensure food security (Zhang et al., 2016) and as functional adaptation scenarios for emerging ecosystems (Catarino et al., 2015). Advances in the field of GMOs would represent the opportunity for the emergence of unique organisms, genetically adapted to situations imposed by anthropic influence, with a high potential for these new individuals to end up being anthropogenically favored due to advances in modern biotechnology, not only limited to plantations, also, cattle and fishing activities (Van Eenennaam, 2017; Forabosco et al., 2013).

Finally, it would be necessary to rethink new models of resource management and human development, and an alternative found that could be applied in the Bio-Anthropogenic Model would be that of "green consumerism", often seen as a form of conscious consumption aimed at the social-environmental perspective (Wiener & Doesher, 1991). The careful choice of products and services would be one of the first steps to achieve less harmful conditions to the environment, mitigating the anthropogenic impact on the environment without significantly compromising the lifestyle of populations similar to Western capitalists (Moisander, 2007). This situation can be illustrated by the creation of the Sustainable Development Goals (SDGs) by the United Nations, which aims to achieve large scales of sustainable and inclusive development across the globe, having implementations of technologies and lifestyle habits considered "nature-friendly" (General Assembly, 2015; Gupta & Vegelin, 2016).

General properties of products and services, classified within the "green consumerism", can be traced by following the criteria of popular consumption guides aimed at environmental preservation (Scheffer, 2013). The "green products" (Elkington et al., 1990) would then be classified by a series of characteristics such as: Not presenting a risk to people's health and the environment, not causing ecological damage during the manufacture and use of the product, consuming quantities of energy considered economical, not causing unnecessary waste due to packaging or programmed obsolescence, in addition to the non-use of material derived from endangered species or ecosystems.

*Wear Out Model*

The last model would be Wear Out, which refers to the continuity of the Evolution of the Human Niche, but with the absence of investments for similar development of green technologies or sustainable and inclusive development (General Assembly, 2015; Gupta & Vegelin, 2016), as well as forms of population occupation with less environmental impacts and global Ecological Restoration techniques. In view of the excessive and damaging exploitation and transformation of resources, as well as the increase in the human occupation rate on the planet (Boivin et al., 2016) to the detriment of the rate occupied by ecosystems themselves, in addition to the generation of massive polluting residues, this model would then be characterized by the wear and tear of several ecosystem services and their by-products (Constanza, 1997), leading to a possible environmental collapse.

When analyzing the pillars of the niche models for *Homo sapiens*, we can see that the characteristics that would stand out, in this scenario, would be our ability to synthesize perceptual pattern contexts through the extremely complex manipulation of the ecology around us, in which we impose selective pressures (Ellis et al., 2016), often creating unsustainable conditions for the existence of several species, would synthesize scenarios of extinction of fauna (Surovell et al., 2005; Johnson, 2006; Koch & Barnosky, 2006) and flora (Johnson , 2009), due to the impacts and disturbances of anthropic character.

Due to the excessive exploitation of tropical forest resources, for example, we would have a global increase in average temperature close to 1 °C (Feddema et al., 2005; Fidell et al., 2006; Bala et al., 2007). In addition, levels of tropical deforestation above 50% provide a decline in the structure and functioning of an ecosystem, as is the case of the Amazon, causing changes in convection processes, cloud formation, rain regimes and exposure to solar radiation, synthesizing a scenario of impact on photosynthesis rates and composition of plant species (Lawrence et al., 2015).

In marine environments, changes in water quality, acidification of the oceans and changes in trophic networks, due to the lack of awareness and use of less impacting techniques by international industries, would stand out in the reasons behind the population falls of several species targeted by the fishing industry (Daskalov, 2002), providing the depletion of so-called coastal ecosystem provision services (Jackson et al., 2001). Thus, possibly bringing about the irreversible global collapse of fisheries resources from the second half of the 21st century (WWF, 2015).

The impacts of the construction of the Human Niche do not apply only to other species, but also to *Homo sapiens*. The impacts on human health, due to the emission of pollutants in marine environments related to the consumption of contaminated food would be expressive (Carbery et al., 2018), in addition to the negative influence of the pesticide industry on food security in several areas of agribusiness (Carneiro, 2015), in which both demonstrated an increase in the cancer mortality rate. Among other consequences of ecological impacts, due to the development of our niche, there is also the limitation of cultural and religious expressions, due to the absence of characteristic local species, making it impossible to use them in rituals, cooking and confection of other products (Kideghesho, 2009; Heise, 2016), as illustrated by the case of the extinction of the species of tuna *Thunnus thynnus*, leading Japanese cuisine to lose several typical dishes (Lemos, 2015).

## CONCLUSIONS

The human niche evolution shows that several pathways are opening within the human relationship with the environment. Several ways of life have already been identified allowing us to predict future scenarios and make up new models within the human lifestyles, aiming at their "uncoupling" to the natural environment, in which even the worst scenarios could be avoided by creating new perspective. The Natural-Technological Model would aim to change the evolutionary direction of the human niche, proposing the resumption of ecological relations originating from diverse ecosystems through active strategies of Ecological Restoration, in addition to the synthesis of new technologies and models of human occupation that would aim significantly mitigate global anthropogenic impacts. The Bio-Anthropogenic Model would have as its main characteristics the viability of emerging ecosystems, considering the symbiotic character of anthropogenic-favored species, alien organisms, genetically modified and hybrids to Homo sapiens, in addition we would have the use of passive restoration for degraded environments and adoption of new perceptions of conscious consumption aiming at the socio-environmental good. Finally, the Wear Out Model, in which we would have an absence of investments for the synthesis of new technologies and models of human occupation, and with the continuity of the Evolution of the Human Niche together with the excessive exploitation of natural resources, that would end up in a global collapse, in which Homo sapiens would suffer from precarious global health, loss of biodiversity, difficulty in growing food products, in addition to great cultural losses associated with several species used in traditional customs. Finally, our "choices" and the "random" will determine our future on earth.